\documentclass[fleqn,usenatbib]{mnras}

\usepackage{newtxtext,newtxmath}
\usepackage{siunitx}
\usepackage{dcolumn}
\newcolumntype{d}[1]{D{.}{.}{#1}}
\usepackage[T1]{fontenc}

\DeclareRobustCommand{\VAN}[3]{#2}
\let\VANthebibliography\thebibliography
\def\thebibliography{\DeclareRobustCommand{\VAN}[3]{##3}\VANthebibliography}

\usepackage{graphicx}	% Including figure files
\usepackage{amsmath}	% Advanced maths commands
\title[Searching for Triple Systems Unbound by Supernovae]{Searching for Triple Systems Unbound by Supernovae}

\author[K Barboza et al.]{
Karina Barboza,$^{1}$\thanks{E-mail: barboza.21@osu.edu}
Christopher S. Kochanek,$^{1, 2}$
\\
$^{1}$Department of Astronomy, The Ohio State University, 140 West 18th Avenue, Columbus OH 43210, USA\\
$^{2}$Center for Cosmology and AstroParticle Physics, The Ohio State University, 191 W. Woodruff Avenue, Columbus OH 43210, USA
}

\date{Accepted XXX. Received YYY; in original form ZZZ}

\pubyear{2023}

\begin{document}
\label{firstpage}
\pagerange{\pageref{firstpage}--\pageref{lastpage}}
\maketitle

\begin{abstract}
A large fraction of massive stars are found in higher order systems where the presence of a tertiary may significantly modify the system's evolution. In particular, it can lead to increased numbers of compact object binaries and accelerate their mergers with important implications for gravitational wave observations. Using Gaia, we constrain the number of Galactic supernovae that produce unbound triples. We do this by searching 8 supernova remnants for stars with consistent Gaia parallaxes and paths intersecting near the center of the supernova remnant at a time consistent with the age of the remnant. We find no candidates for unbound triple systems. Combined with prior work, less than 11.4\% of supernovae leave behind unbound triples at a 90\% confidence limit. The absence of such systems limits their role in the evolution of massive stars and the formation of merging compact objects.
\end{abstract}

\begin{keywords}
supernova remnants -- triples -- compact objects
\end{keywords}

%%%%%%%%%%%%%%%%%%%%%%%%%%%%%%%%%%%%%%%%%%%%%%%%%%

%%%%%%%%%%%%%%%%% BODY OF PAPER %%%%%%%%%%%%%%%%%%

\section{Introduction}
Most massive stars have at least one stellar companion \citep[e.g.,][]{Evans_2006, Kobulnicky_2007, Sana_2014, Moe_2017} and are expected to interact at some point during their evolution \citep[][]{Sana_2012}. For massive stars, triple systems are likely more common than binaries \citep[][]{Moe_2017}. While these surveys have provided valuable insights into the multiplicity of massive stars, the path to mergers or interacting compact objects is long and intricate. To understand the expected properties of supernovae (SN) and compact objects, we must understand the evolution of such systems. 

Binary population synthesis (BPS) models are used to try to understand and predict the evolution of binary populations \citep[e.g.,][]{Belczynski_2008, Eldridge_2017}. Recent expansions in BPS studies now encompass the tracking of triple or quadruple systems with their significantly richer dynamical properties \citep[e.g.,][]{Toonen_2020, Hamers_2021}. The introduction of a tertiary to a system's evolution can lead to enhanced tidal interactions \citep[][]{Fabrycky_2007}, mass transfer in eccentric orbits \citep[][]{Thompson_2011, Shappee_2013, Toonen_2020},  increase the number of compact object binaries \citep[][]{Thompson_2011, Shappee_2013} and accelerate their mergers \citep[][]{Perets_2009}. These accelerated mergers can then produce transients such as gamma-ray bursts, Type Ia supernovae (SNe), and gravitational wave mergers \citep[][]{Wen_2003, Perets_2009, Thompson_2011, katz2012rate, Hamers_2013, Kimpson_2016, Portegies_Zwart_2016, Abbott_2021}. 

Aside from the standard issues in stellar evolution, there are complexities in BPS models such as mass transfer, common envelope evolution, neutron star and black hole natal kicks \citep[see, e.g., the review of BPS models by][]{Han_2020}, and understanding the connections between a core collapse and its outcomes \citep[supernovae, failed supernovae, neutron stars, or black holes see, e.g.,][]{Pejcha_2015, Kochanek_2014, Sukhbold_2016}. These physical effects have significant uncertainties, but they are crucial to predictions, especially for the population of merging compact objects. One way to reduce the uncertainties is to determine the statistics of binary and higher order systems at the time of stellar death rather than birth. This can be done through studies of Galactic SNRs.

For SNRs with known compact objects, finding bound systems is relatively straight forward, as the companion and the remnant are co-located and, in the three known cases, interacting \citep[SS433; HESS J0632$+$057 in MWC 148, \citealt{Hinton_2009}; and 1FGL J1018.6$-$5856 in 2MASS J10185560$-$5856489, \citealt{Corbet_2011}] {Hillwig_2008}. \citet{Kochanek_2019} searched for surviving but non-interacting binaries and found none in 23 SNRs. 

Detecting unbound binaries or higher order systems in supernova remnants (SNRs) is challenging because the remnant and the former companion are not co-located. This can become even more challenging in the case where there is no identified remnant, as illustrated by the many studies searching for a former stellar companion to the Type Ia SN Tycho \citep[e.g.,][]{Ruiz_Lapuente_2004, Kerzendorf_2018, Ihara_2007}. Typically, the former companion should remain close to the center of the SNR. However, the distortion and asymmetry of many SNRs makes it challenging to accurately determine the search region from the structure of the remnant. \citet{Boubert_2017} and \citet{Lux_2021} have attempted such searches for core-collapse SNRs, but this approach likely has a high false positive rate \citep[see][]{Kochanek_2021}.

In the case where a remnant is identified and the SNR is very young, the process becomes straightforward because a surviving companion must be nearly co-located with the remnant, as in the cases of the Crab, Cas A, and SN 1987A \citep[]{Kochanek_2018}. If the remnant has a measured proper motion (and preferably a parallax), this can be combined with stellar proper motions and parallaxes to search for companions whose back projected trajectories intersect with that of the remnant near the center of the SNR at a reasonable age. This was done by \citet{Din_el_2015}, and expanded by \citet{Fraser_2019}, \citet{Kochanek_2021}, and \citet{Kochanek_2023}. To date, there is only one good candidate for an unbound binary companion, 
HD~37424 in S147 (\citealt{Din_el_2015}).  The present statistics are that 77\% (58-90\% at 90\% confidence) are not binaries
at death, 9\% (4-19\%) are bound binaries, 6\% (2-13\%) are interacting binaries, $<9\%$ are non-interacting binaries and 12\% (3-31\%) are unbound binaries (\citealt{Kochanek_2023}).

Finding unbound triples is actually easier than finding unbound binaries because it is not dependent on even identifying the remnant, let alone having its proper motions or parallaxes. One can simply look for a pair of stars with consistent parallaxes whose back projected trajectories intersect with the remnant near the center of the SNR. This was first done by \citet{Kochanek_2021}, who  studied 10 SNRs and found no candidates, which implied that fewer than 18.9\% core-collapse SN lead to unbound triples at 90\% confidence. Here we expand this search for unbound triples to 8 additional SNRs using Gaia DR3. We outline the method in \S \ref{sec:method}, discuss the 8 SNRs and the search results in \S \ref{sec:sample} and summarize them in \S \ref{sec:summary}.

\section{Identifying Stellar Companions} \label{sec:method}

For each SNR, we search Gaia DR3 \citep[]{gaia_2023, gaia_2016} for stars within a radius $R/R_{SNR} < 1$ where $R_{SNR}$ is the radius of the SNR and with a magnitude limit $G_{lim}$ chosen to reach stars with $ M \gtrsim M_{\odot}$ given the distance and extinction to the SNR. We discuss the adopted distances and extinctions in \S \ref{sec:sample}. We identify candidate triples using the methods from \citet{Kochanek_2021}, outlined in this section and summarized in Table \ref{tab:steps}. While all the SNRs contain neutron stars, we primarily search for possible unbound pairs of stars.

First we restrict our candidates to those that can intersect within the neutron star's path in the time frame $-t_m < t < 0$ for a neutron star velocity of $|v_{ns}| < v_{max}$. The positional difference between the neutron star and the star can be written as $\Delta\overrightarrow{\text{x}} = d\Delta\overrightarrow{\alpha}$ for a distance $d$ where $\Delta\overrightarrow{\alpha} = \overrightarrow{\alpha}_{ns} - \overrightarrow{\alpha}_*$, the velocity of the  star is $\overrightarrow{v}_{*} = d\overrightarrow{\mu}_{*}$, so the neutron star velocity ($v_{ns}$) needed to intercept the compact object at time t is

\begin{equation}
\centering
   % \overrightarrow{v}_{ns} = (v^2_*(t) - v^2_{ns} -\Delta x^2 t^{-2})/(2t^{-1})
   v^2_{ns} = v^2_*+2t^{-1} \overrightarrow{v}_* \cdot \Delta \overrightarrow{x}+\Delta x^2 t^{-2}.
  % v^2_*(t)= v^2_{ns}+2t^{-1} \overrightarrow{v}_{ns} \cdot \Delta \overrightarrow{x}+\Delta x^2 t^{-2}
\label{eq:vstar}
\end{equation}
The angular coordinate differences incorporate the necessary cos$\delta$ term for the separation in RA and the angular separations are small enough to just work in the tangent plane. We keep the stars which can intercept the remnant at $R<R_{SNR}/4$ with $|v_{ns}|  \leq v_{max} = 400$ km/s and $-10^5 \leq t \leq 0$ years.

Next we search for pairs of these stars with consistent parallax estimates. If the parallaxes are $\varpi_i \pm \sigma_{\varpi, i} $ where $i = 1,2$, then their joint parallax is

\begin{equation}
\centering
    \varpi = \left[\sum_i \frac{\varpi_i}{\sigma^2_{\varpi, i}} \right] \left[\sum_i \frac{1}{\sigma^2_{\varpi, i}} \right]^{-1}
\label{eq:parallax}
\end{equation}
with a goodness of fit

\begin{equation} 
\centering 
    \chi^2_{\varpi} = \sum_i \left(\frac{\varpi_i-\varpi}{\sigma^2_{\varpi, i}} \right)^2.
\label{eq:chi2}
\end{equation}
Pairs meeting the requirement $\chi^2_{\varpi} < 9$ are kept as potential candidates for unbound triple systems. 

The paths of a pair of unbound stars need not precisely intersect at a particular time because the disrupted binaries had a finite size. We introduce a maximum binary separation $a_{max}$ which corresponds to an angle of 

\begin{equation}
\centering
   \sigma_a=10\farcs0 \left(\frac{a_{max}}{10^4 \text{AU}}\right) \left(\frac{\varpi}{\text{mas}}\right)
\label{eq:amax}
\end{equation}
We must also account for the uncertainties in the proper motions. We do this by treating the true proper motions as an intermediate variable. For example, for the motion in RA we can let the current separation of two objects be $\Delta \alpha_{12} = \alpha_1 - \alpha_2$, their measured current motions be $\mu_{\alpha,i}$, and their true proper motions be $\mu^T_{\alpha,i}$. The cosine declination terms on $\Delta \alpha$ are again implicit. The measured proper motions impose constraints on the true proper motion, leading to the fit statistic

\begin{align}
\label{eq:fitstats}
   \chi^2_{\alpha,12} = \frac{(\mu_{\alpha,1}-\mu^T_{\alpha,1})^2}{\sigma^2_{\alpha,1}} + \frac{(\mu_{\alpha,2}-\mu^T_{\alpha,2})^2}{\sigma^2_{\alpha,2}} + \\
   \frac{\left(\Delta \alpha_{12} + t\left(\mu^T_{\alpha,1}-\mu^T_{\alpha,2}\right)\right)^2}{\sigma^2_\alpha} \nonumber .
\end{align}
Here the first two terms use the measured proper motions to constrain the true proper motions, while the last term constrains the closest approach distance relative to the maximum binary size based on the true proper motions. We can then optimize the fit statistic for the unknown true proper motions to obtain

\begin{align}
\label{eq:optfitstat}
   \chi^2_{\alpha,12} = \frac{(\Delta \alpha_{12}/t + \Delta \mu_{\alpha,12})^2}{\sigma^2_{\alpha,12}}
\end{align}
where $\Delta \mu_{\alpha,12} = \mu_{\alpha,1} - \mu_{\alpha,2}$, and

\begin{align}
\label{eq:sigma2}
   \sigma^2_{\alpha,12} = \sigma^2_{\alpha,1} + \sigma^2_{\alpha,2} + \sigma^2_{\alpha}/t^2.
\end{align}
We now define $\Delta_{\delta,12} = \Delta \delta_{12}/t+\Delta \mu_{\delta,12}$ to obtain

\begin{align}
\label{eq:chimu2}
   \chi^2_\mu = \frac{\Delta^2_{\alpha,12}}{\sigma^2_{\alpha,12}} + \frac{\Delta^2_{\delta,12}}{\sigma^2_{\delta,12}},
\end{align}
with the declination term. A value of $\chi^2_\mu \simeq 3$ roughly corresponds to the stars passing within a distance $a_{max}$ of each other with proper motions that are each within $1\sigma$ of their measured values. 

Because of equation \eqref{eq:sigma2}, the optimization of $\chi^2_\mu$ is non-linear in $t^{-1}$. To solve this problem, we keep the time in equation \eqref{eq:sigma2} fixed to determine a new optimal time, use this new time to update the equation, and repeat the process to convergence. The positional uncertainties are excluded because they are unimportant, but they could be incorporated as an error term such as $\sigma_\alpha$.

\begin{figure}
\centering
\includegraphics[scale=0.36]{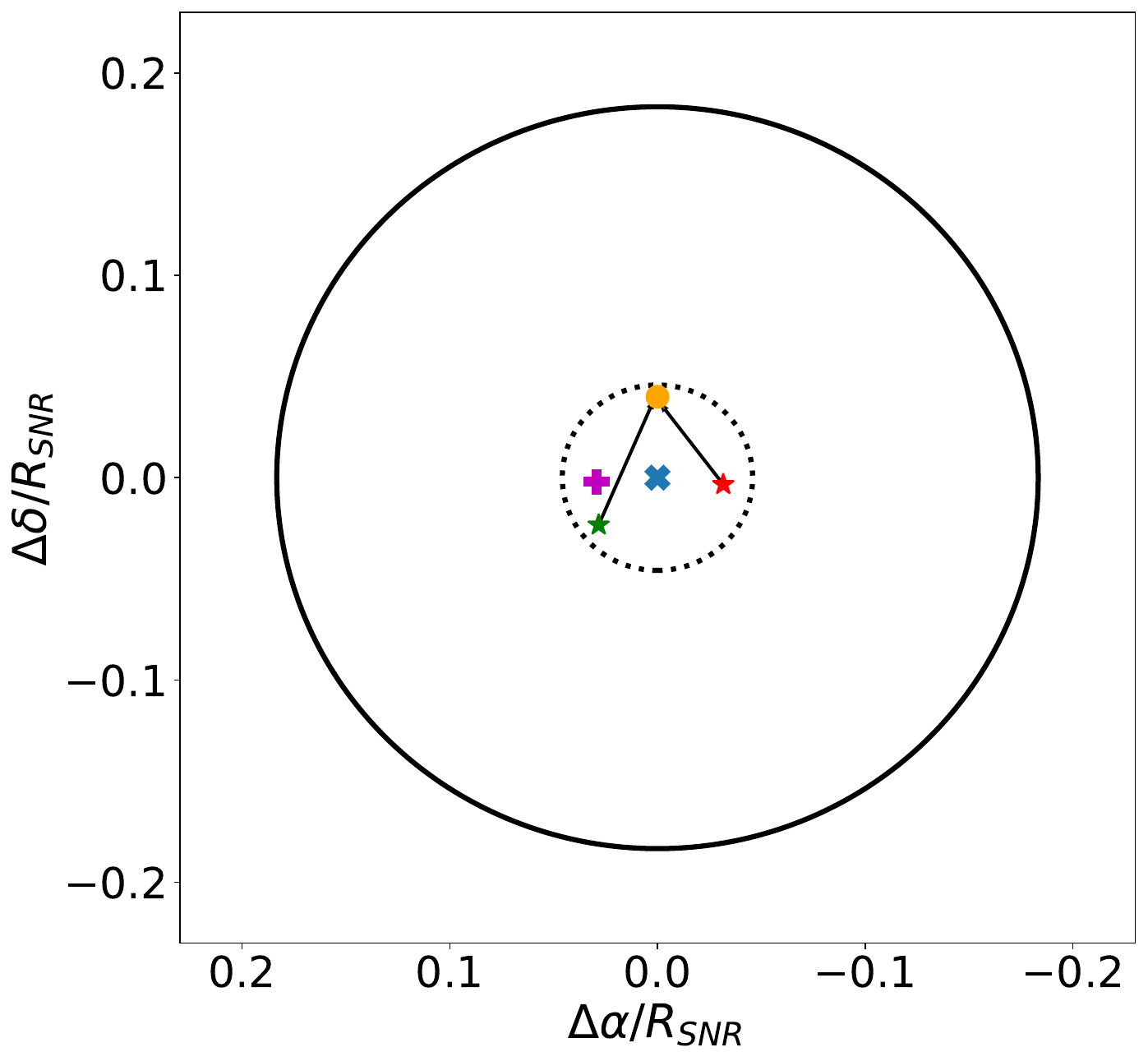}
\caption{An example of a pair which could be a good candidate based on the geometry of the SNR G$065.7+01.2$. The current positions of the stars are given by the red and green stars and the position at the optimal intersection time is given by the orange circle. The blue cross mark is the estimated center of the SNR, the magenta plus sign is the location of the remnant, and the dotted circle is the $R_{SNR}/4$ region we require for an acceptable candidate. Unfortunately, we found no candidates which met all the requirements.}

\label{fig:intersxn}
\end{figure}

With the uncertainties fixed, the closest approach time is

\begin{equation}
\label{eq:tm}
   t_m = -\frac{\Delta\alpha_{12}^2\sigma_{\delta,12}^2 +\Delta\delta_{12}^2\sigma_{\alpha,12}^2}{\Delta\alpha_{12}\Delta\mu_{\alpha,12}\sigma_{\delta,12}^2 + \Delta\delta_{12}\Delta\mu_{\delta,12}\sigma_{\alpha,12}^2} 
\end{equation}
with a goodness of fit of

\begin{align}
\label{eq:chi12}
   \chi_{12}^2 = \frac{(\Delta\delta_{12}\Delta\mu_{\alpha,12} - \Delta\alpha_{12}\Delta\mu_{\delta,12})^2}{\Delta\alpha_{12}^2\sigma_{\delta,12}^2+ \Delta\delta_{12}^2\sigma_{\alpha,12}^2} .
\end{align}
We require $\chi^2_{\mu} <9$, an intercept time of $-10^5 \leq t_m \leq 0$ years and an intersection point whose radius relative to the SNR center satisfies $R/R_{SNR} < 1/4$. We can then check the consistency of the candidates with the estimated age and distance of the SNR. Figure \ref{fig:intersxn} illustrates a model of a good candidate. The dashed circle is the SNR search radius $R_{SNR}/4$, the red and green stars are the current positions of the stars, and the orange circle is the intersection point. This is a model candidate because we find no pairs meeting all the selection criteria.

We employed PARSEC isochrones from \citet{Pastorelli_2020} to categorize the stars into different mass ranges, based on their positions on an extinction-corrected $M_G$ versus $B_P - R_P$ color-magnitude diagram. This is the only aspect of the calculation involving distances, and employing the inverse parallax is sufficiently significantly accurate because of the steepness of stellar mass-luminosity relations. We grouped stars by applying limits of 1, 2, 3, 4 and 5$M_{\odot}$, with the color-dependent absolute magnitude limits given in Table \ref{tab:cmd}. For example a star with a $B_P - R_P <0$ and  $M_G < -0.4$ should have a mass $>5M_{\odot}$. 

\section{Sample and Results} 
\label{sec:sample}
From the 23 SNRs in Table 2 of \citet{Kochanek_2019}, we examined eight: G065.7$+$01.2, G078.2$+$02.1, G106.3$+$02.7, G114.3$+$00.3, G119.5$+$10.2, G189.1$+$03.0, G266.2$-$01.2, G291.0$-$00.1. The positions and radii of these SNRs are shown in Table \ref{tab:snrcands}. Of the remaining 15 remnants, 12 were previously studied in \citet{Kochanek_2021} and \citet{Kochanek_2017}, and the remaining 3 (G034.7$-$00.4, G320.4$-$01.2, G332.4$-$00.4) required using such faint stars that the searches were dominated by false positives. We adopt the SNR centers and diameter from \citet{Green_2019}. For the stellar extinctions we used {\tt mwdust} \citep[][]{Bovy_2016}, which is based on the which combine the \citet{Drimmel_2003}, \citet{Marshall_2006}, and \citet{Green_2019} dust models. These models generally agree well with X-ray estimates of the column density N(H) for the distance of the SNR.

The selection process is outlined in Table \ref{tab:steps}. For the first step, we queried GAIA DR3 \citet{gaia_2023} to find all the stars within the radius of the SNR where the limiting magnitude $G_{lim}$ is chosen based on the distance and extinction of the remnant. From these stars, in step 2 we selected stars whose paths intersect the neutron star with $R/R_{SNR}< 1/4, |v_{ns}|< 400$ km/s and $-10^5 \leq t \leq 0$ years (equation \ref{eq:vstar}). As shown in Table \ref{tab:steps}, this is where we lost most of the initial sample. Next in step 3, we determined which pairs of stars had consistent parallaxes $\chi_{\varpi}^2 <9$ (equations \ref{eq:parallax} and \ref{eq:chi2}). For step 4, we determined which of these pairs of stars could intercept with $\chi_{\mu}^2 <9$ (equation \ref{eq:chimu2}). Finally, our last constraint was that star pairs intersect within $R_{SNR}/4$ and at a time between $-10^5 < t_m < 0$ years, which none of our candidates satisfied. 

As we tested the search, we discovered two pairs which had present day separations of $2.6 \times 10^4$ AU and $4.9 \times10^5$ AU with statistically consistent parallaxes and proper motions. These are almost certainly wide binaries. We then searched all the pairs for cases which have present day separations $<10^6$ AU and statistically consistent parallaxes and proper motions. We defined statistically consistent as  $\chi^2_b = \chi^2_\varpi+\chi^2_{\mu,\alpha}+\chi^2_{\mu,\delta}<9$ where $\chi^2_\varpi$ is simply Eqn \ref{eq:chi2} and $\chi^2_{\mu,\alpha}$ and $\chi^2_{\mu,\delta}$ are the analogous fit statistics for the two components of the proper motions. Of all pairs found in step 3, 11 have a present day separation $<10^6$ AU with $\chi^2_b < 9$. 

SNR G065.7+01.2 contains the resolved pulsar wind nebula (PWN) DA 495 and central object J1952.2+2925 \citep[]{Arzoumanian_2004}. \citet{Kothes_2008} calculate the age of the remnant to be $\sim 20,000$ yrs. The first star pair candidate is located at 3.3 kpc, a distance consistent with that of the remnant. The distance to the second star pair candidate is 6.5 kpc which is not consistent with the remnant. While these two candidates meet the two $\chi^2$ requirements, neither candidate intersects on a reasonable time scale or close to the remnant's center. Both of these star pair candidates are almost certainly wide binaries since they have  present day separations of $3.1 \times 10^5$ AU and $4.9 \times 10^5$ AU respectively, and statistically consistent parallaxes and proper motions. 

The SNR G078.2+02.1 is associated with the $\gamma$-ray pulsar PSR J2021+4026 \citep[]{Abdo_2009}. \citet{Trepl_2010} estimated an age of 6.6 kyr which is in agreement with the age from \citet{Mavromatakis_2003} and an order of magnitude less than the $\tau \sim 77$ kyr spin-down age of the pulsar \citep[]{Trepl_2010}. Of the 21 candidates found in step 2, none met the two $\chi^2$ requirements. 

SNR G106.3+02.7 contains the radio pulsar PSR J2229+6114. \citet{Halpern_2001b} determine the characteristic age of the pulsar as 10,460 years. None of the 11 candidates from step 2 met the two $\chi^2$ requirements.

The remnant G114.3+00.3 is associated with radio pulsar PSR B2334+61 \citep[]{Dewey_1985}.\citet{Cioff_1988} estimate the age of the remnant to be 7700 years. None of the stars from our initial search made it past step 2.

G119.5+10.2 (CTA 1) contains the X-ray source RX J0007.0+7303 \citep[]{Halpern_2004}, later found to be a $\gamma$-ray \citep[]{Abdo_2009} and X-ray pulsar \citep[]{Lin_2010}. The pulsar has a spin-down age of $ \tau_c \sim 14,000$ yrs which is consistent with the age of the remnant \citep[10,000–20,000 yrs,][]{Lin_2010}. None of the 3 candidates from step 2 met the two $\chi^2$ requirements.

The SNR G189.1+03.0 (IC 443) contains the neutron star CXO J061705.3+222127 and the PWN G189.22+2.90 \citep[]{Keohane_1997}. \citet{Chevalier_1999} estimate an age of $\sim 30,000$ years for the remnant. None of the 4 candidates from step 2 met the two $\chi^2$ requirements.

SNR G266.2-01.2 (Vela Jr.) contains the X-ray source CXOU J085201.4-461753 \citep[]{Pavlov_2001}. The characteristic age of the remnant is 140 kyr according to \citet{Kramer_2003} while \citet{Allen_2014} estimate an age of 2.4-5.1 kyr. The one candidate left after step 2 did not meet the two $\chi^2$ requirements.

G291.0-00.1 is associated with the X-ray pulsar CXOU J111148.6-603926 \citep[]{Slane_2012}. The age of this remnant is estimated to be 1300 years \citep[]{Harrus_1998}. None of the stars from our initial search made it past step 2.

\section{Conclusions}\label{sec:summary}
We searched for unbound triple systems in the 8 SNRs listed in Table \ref{tab:snrcands} and found none. Combined with the 10 systems considered by \citet{Kochanek_2021}, we have now searched M = 18 SNRs and found no good candidates for unbound triples. This limits the fraction $f_T$ of triple systems becoming unbound at stellar death to the binomial probability
\begin{equation}
\centering
   P(<f_T) = 1-(1-f_T)^{1+M}. 
\label{eq:Binomial_prob}
\end{equation}
Thus, less than $f_T < 11.4$\% of SNe leave behind unbound massive triples at a 90\% confidence limit. This limit along with the measurements of bound and unbound binary fractions from \citet{Kochanek_2019}, \citet{Kochanek_2021}, and \citet{Kochanek_2023} provide constraints on multiplicity that should inform and test BPS models. Recently, \citet{Burdge2024} found that the black hole low mass X-ray binary V404~Cygni was likely a triple system, and discuss the challenges of keeping such a system bound.  We note that these limits on binary systems effectively apply to the abundance of surviving triple systems as well.

\section*{Acknowledgements}
CSK is supported by the NSF grants AST-1908570 and AST-2307385. KB is supported by the Heising-Simons Foundation through grant number 2022-3533. We are thankful to be able to think and work on the ancestral and contemporary territory of the Shawnee, Potawatomi, Delaware, Miami, Peoria, Seneca, Wyandotte, Ojibwe, and many other Indigenous peoples. 

%%%%%%%%%%%%%%%%%%%%%%%%%%%%%%%%%%%%%%%%%%%%%%%%%%
\section*{Data Availability}
 
All data used in this paper are publically available.

\begin{table*}
	\centering
	\caption{SNR candidates and their properties: The column densities N(H) are, in order, from \citet{Arzoumanian_2008}, \citet{Hui_2015}, \citet{Halpern_2001a}, \citet{McGowan_2006}, \citet{Slane_1997}, \citet{Gaensler_2006}, \citet{Pavlov_2001}, \citet{Slane_2012}. Distance estimates are from \citet{Karpova_2016}, \citet{landecker_1980}, \citet{Halpern_2001b}, \citet{McGowan_2006}, \citet{Pineault_1993}, \citet{Fesen_1984}, \citet{Liseau_1992}, \citet{Slane_2012}.}
	\label{tab:snrcands}
    \makebox[-1cm][c]{   
    \resizebox{0.83 \textwidth}{!}{
    \begin{tabular}{cccccSc}
         \hline
         \multicolumn{1}{c}{\textbf{SNR}}&
         \multicolumn{1}{c}{\textbf{Center}} &
         \multicolumn{1}{c}{\textbf{$R_{SNR}$}} &
         \multicolumn{1}{c}{\textbf{d}} &
         \multicolumn{1}{c}{\textbf{N(H)}} & 
         \multicolumn{1}{c}{\textbf{$G_{lim}$}} &
         \multicolumn{1}{c}{\textbf{Remnant}} \\
         \multicolumn{1}{c}{} &
         \multicolumn{1}{c}{} &
         \multicolumn{1}{c}{(')} &
         \multicolumn{1}{c}{(kpc)} &
         \multicolumn{1}{c}{($cm^{-2}$)} &
         \multicolumn{1}{c}{(mag)} &
         \multicolumn{1}{c}{} \\
         \hline 
            G$065.7+01.2$ & 19:52:10 $+$29:26:00 & $11$ & $3.0 \pm 2.0$ & $(2-7) \times 10^{21}$ & 15.5 & PSR J$1952.2+2925$\\ 
            G$078.2+02.1$ & 20:20:50 $+$40:26:00 & $30$ & $1.5 \pm 0.5$ & $6.4^{+0.8}_{-1.8} \times 10^{21}$ & 15 & PSR J$2021+4026$\\
            G$106.3+02.7$ & 22:27:30 $+$60:50:00 & $19$ & 3.0 & $6.3 \pm 1.3 \times 10^{21}$ & 16 & PSR J$2229+6114$\\
            G$114.3+00.3$ & 23:37:00 $+$61:55:00 & $35$ & $3.2 \pm 1.7$ & $(0.26 - 1.0)\times 10^{22}$ & 17 & PSR B$2334+61$ \\
            G$119.5+10.2$ & 00:06:40 $+$72:45:00 & $45$ & $1.4 \pm 0.3$ & $2.8^{+0.6}_{-0.5} \times  10^{21}$ & 12 & RX J$0007.0+7303$\\ 
            G$189.1+03.0$ & 06:17:00 $+$22:34:00 & $23$ & $1.7 \pm 0.3$ & $(7.2 \pm 0.6) \times 10^{21}$ & 14.5 & PWN G$189.22+2.90$\\
            G$266.2-01.2$ & 08:52:00 $-$46:20:00 & $60$ & $0.7 \pm 0.2$ & $(3.0 \pm 1.0)\times 10^{21}$ & 11 & CXOU J$085201.4-461753$\\
            G$291.0-00.1$ & 11:11:54 $-$60:38:00 & $7$ & $3.5-5.0$ & $(6.7 \pm 0.7)\times 10^{21}$ & 16.5 & CXOU J$111148.6-603926$\\
         \hline
    \end{tabular}
 
    }
}
\end{table*}

\begin{table*}
  \centering
  \caption{CMD Selection Limits: Mag gives the upper limit on $M_G$ for stars above the listed mass.}
  \begin{tabular}{lcccccccc}
  \hline
  \multicolumn{1}{c}{Mass} &
  \multicolumn{1}{c}{\textbf{Color}} &
  \multicolumn{1}{c}{Mag} &
  \multicolumn{1}{c}{\textbf{Color}} &
  \multicolumn{1}{c}{Mag} &
  \multicolumn{1}{c}{Color} &
  \multicolumn{1}{c}{Mag} &
  \multicolumn{1}{c}{Color} &
  \multicolumn{1}{c}{Mag} \\
  \hline
  \multicolumn{1}{c}{$>1M_\odot$} &
  \multicolumn{1}{c}{$<1.5$} &
  \multicolumn{1}{c}{$\hphantom{-}5.5$} &
  &
  &
  \multicolumn{1}{c}{$1.5$-$2.0$} &
  \multicolumn{1}{c}{$-1-3(B_P-R_P-1)$} &
  \multicolumn{1}{c}{$>2$} &
  \multicolumn{1}{c}{$-4$} \\
  \multicolumn{1}{c}{$>2M_\odot$} &
  \multicolumn{1}{c}{$<0.7$} &
  \multicolumn{1}{c}{$\hphantom{-}2.0$} &
  \multicolumn{1}{c}{$0.7$-$1.1$} &
  \multicolumn{1}{c}{$\hphantom{-}3.0$} &
  \multicolumn{1}{c}{$1.1$-$2.0$} &
  \multicolumn{1}{c}{$-1-3(B_P-R_P-1)$} &
  \multicolumn{1}{c}{$>2$} &
  \multicolumn{1}{c}{$-4$} \\
  \multicolumn{1}{c}{$>3M_\odot$} &
  \multicolumn{1}{c}{$<0.7$} &
  \multicolumn{1}{c}{$\hphantom{-}1.0$} &
  \multicolumn{1}{c}{$0.7$-$1.1$} &
  \multicolumn{1}{c}{$\hphantom{-}2.0$} &
  \multicolumn{1}{c}{$1.1$-$2.0$} &
  \multicolumn{1}{c}{$-1-3(B_P-R_P-1)$} &
  \multicolumn{1}{c}{$>2$} &
  \multicolumn{1}{c}{$-4$} \\
  \multicolumn{1}{c}{$>4M_\odot$} &
  \multicolumn{1}{c}{$<0.0$} &
  \multicolumn{1}{c}{$\hphantom{-}0.5$} &
  \multicolumn{1}{c}{$0.0$-$1.3$} &
  \multicolumn{1}{c}{$-1.9$} &
  \multicolumn{1}{c}{$1.3$-$2.0$} &
  \multicolumn{1}{c}{$-1-3(B_P-R_P-1)$} &
  \multicolumn{1}{c}{$>2$} &
  \multicolumn{1}{c}{$-4$} \\
  \multicolumn{1}{c}{$>5M_\odot$} &
  \multicolumn{1}{c}{$<0.0$} &
  \multicolumn{1}{c}{$-0.4$} &
  \multicolumn{1}{c}{$0.0$-$1.3$} &
  \multicolumn{1}{c}{$-1.9$} &
  \multicolumn{1}{c}{$1.3$-$2.0$} &
  \multicolumn{1}{c}{$-1-3(B_P-R_P-1)$} &
  \multicolumn{1}{c}{$>2$} &
  \multicolumn{1}{c}{$-4$} \\
  \hline
\multicolumn{9}{l} {
  Mag gives the upper limit on $M_G$ for each $B_P-R_p$ color range. }
  \end{tabular}
  \label{tab:cmd}
\end{table*}

\begin{table*}
    \caption{Star pair candidates during the selection process.}
    \label{tab:steps}
    \makebox[1\textwidth][c]{       %centering table
    \resizebox{0.95 \textwidth}{!}{
    \begin{tabular}{ lllcccccccc } 
        \hline
        \multicolumn{3}{c}{Steps}&
        \multicolumn{1}{c}{G$065.7+01.2$}&
        \multicolumn{1}{c}{G$078.2+02.1$}&
        \multicolumn{1}{c}{G$106.3+02.7$}&
        \multicolumn{1}{c}{G$114.3+00.3$}&
        \multicolumn{1}{c}{G$119.5+10.2$}&
        \multicolumn{1}{c}{G$189.1+03.0$}&
        \multicolumn{1}{c}{G$266.2-01.2$}&
        \multicolumn{1}{c}{G$291.0+00.1$}\\
        \hline
        1 &initial stars & stars & 700 & 554 & 392 & 6336 & 132 & 252 & 88 & 1090 \\
        2 &intersect NS & stars & 92 & 21 & 11 & 0 & 3 & 4 & 1 & 0 \\
        3 &consistent $\varpi$& pairs & 62 & 0 & 0 & 0 & 0 & 0 & 0 & 0 \\
        4 &intersect & pairs & 2 & 0 & 0 & 0 & 0 & 0 & 0 & 0 \\
        5 &time/ radius & pairs & 0 & 0 & 0 & 0 & 0 & 0 & 0 & 0 \\
        \hline
    \end{tabular}
    }
}
\end{table*}


\bibliographystyle{mnras}
\bibliography{newbib} % if your bibtex file is called example.bib

%%%%%%%%%%%%%%%%%%%%%%%%%%%%%%%%%%%%%%%%%%%%%%%%%%

%%%%%%%%%%%%%%%%% APPENDICES %%%%%%%%%%%%%%%%%%%%%

%%%%%%%%%%%%%%%%%%%%%%%%%%%%%%%%%%%%%%%%%%%%%%%%%%

% Don't change these lines
\bsp	% typesetting comment
\label{lastpage}
\end{document}